\documentclass[epj]{svjour}
\usepackage{graphics}
\usepackage{amsmath}
\usepackage{amsfonts}
\usepackage{amssymb}
\usepackage[normalem]{ulem}

\begin{document}
\title{Definition of frame-invariant Soret coefficients\\for ternary mixtures}
\author{Jos\'e M. Ortiz de Z\'arate
}                     
%
%
\institute{Facultad de Ciencias F\'{\i}sicas. Universidad Complutense. Plaza de las Ciencias, 1. 28040 Madrid, Spain}
\date{Received: date / Revised version: date}
%
\abstract{The definition of the Soret coefficient of a binary mixture includes a concentration prefactor, $x(1-x)$ when mol fraction $x$ is used, or $w(1-w)$ when mass fraction $w$ is used. In this paper the physical reasons behind this choice are reviewed, emphasizing that the use of these prefactors makes the Soret coefficient invariant upon change in the reference frame, either mass or molar. Then, it is shown how this invariance property can be extended to ternary mixtures by using an appropriate concentration prefactor in matrix form. The work is completed with some considerations of general non-isothermal diffusion fluxes, binary limits of the concentration triangle, selection of the dependent concentration in a ternary mixture, use of molar concentrations and, finally, extension to multi-component mixtures.
} 
\maketitle
\section{Introduction\label{S1}}

Thermodiffusion is a transport phenomenon that generically exists in non-isothermal multicomponent mixtures. The presence of a temperature gradient $\boldsymbol{\nabla}{T}$ induces not only a heat flow, but also a contribution to the diffusion (mass) flux. In the theoretical framework of linear non-equilibrium thermodynamics~\cite{DeGrootMazur,Demirel,LebonJouCasas,KjelstrupBook2010} the thermodiffusion flow is, in a first approximation, proportional to $\boldsymbol{\nabla}{T}$.

When a steady temperature gradient is applied to a multi-component liquid mixture thermodiffusion induces a separation of its components, so that concentration gradients develop in the system. In the case of a binary liquid mixture and a stationary temperature gradient, after a transient, a non-equilibrium steady state is reached where a constant (in time) concentration gradient is established. In isotropic fluids\footnote{The contains of this paper  only refer to isotropic fluids} the applied temperature gradient and the induced concentration gradient are always parallel (or antiparallel). To quantify thermodiffusion in binary mixtures, because of pioneering research by Charles Soret (1879, see~\cite{PlattenCosteseque}), the so-called Soret coefficient $S_T$ (units of K$^{-1}$) is defined as proportional to the ratio of these steady concentration and temperature gradients, namely,
\begin{equation}\label{E01}
x (1-x) S_T~\nabla{T}=-\nabla{x},
\end{equation}
where $x$ is the (average) concentration of the mixture in mol fraction. For the purpose of this paper it is important to clearly distinguish between the various possible ways of specifying the composition of a mixture, thus, symbol $x$ shall be used to denote mol fraction, $w$ for mass fraction (both dimensionless) and $c$ for molar concentration (units of $\text{mol}~\text{m}^{-3}$). The last will be, in particular, considered later, in Sect.~\ref{S5}.

Note the presence of concentration prefactor $x(1-x)$ in the definition~\eqref{E01} of the Soret coefficient of a binary mixture. There are various reasons for introducing this prefactor:

1. The prefactor forces the thermodif\-fusion-induced concentration gradient to vanish in the two pure component limits, \emph{i.e.}, for $x\to0$ and $x\to1$. In this regard, the prefactor is expected to carry most of the concentration dependence of thermodiffusion, so that the Soret coefficient $S_T$ will only depend weakly on concentration. Indeed, in many cases, a linear dependence on $x$ is enough to represent $S_T$ well over the entire range of concentration, $x\in[0,1]$.

2. Most importantly, the prefactor makes the $S_T$ value invariant under change in the representation of concentration. Indeed, since
\begin{equation}\label{E02}
\frac{\nabla{x}}{x(1-x)}=\frac{\nabla{w}}{w(1-w)},
\end{equation}
the numerical value experimentally measured for $S_T$ will be identically the same, independently of whether it is computed like in Eq.~\eqref{E01} with concentrations in mol fraction $x$, or by
\begin{equation}\label{E03}
w (1-w) S_T~\nabla{T}=-\nabla{w},
\end{equation}
with concentrations in mass fraction $w$.

Notice that property \#2 above is quite convenient from both a practical and an experimental point of view. It makes life easier for scientists obtaining $S_T$ values in binary mixtures, either experimentally or by computer simulations. They can continue to use their favorite concentration representation, ones in mol fraction~\cite{BeltonTyrrell,WiegandKitaNing,CabreraEtAl,Hafskjold17,DiLecce17} and others in mass fraction~\cite{KohlerMuller95,MialdunShevtsova,FurtadoEtAl}, while comparison between values obtained by different teams is direct, not requiring any conversion or number manipulation.

Historically, the concentration prefactor in its current form, Eqs.~\eqref{E02} or~\eqref{E03}, first appeared in the Enskog formula (1917) for the separation of isotopes by thermal diffusion. It was later adopted for arbitrary binary mixtures, in particular\footnote{I owe this information to a  personal communication by Aliaksandr Mialdun.} through 1942 work by Sybren De Groot~\cite{DeGroot1942}. Interestingly, as reason for introducing $x(1-x)$, de Groot quotes yet another nice property, namely, invariance of $S_T$ under permutation of components (it simply changes sign, see Sect.~\ref{S3}). Other authors at that time were using proportionality to concentration, which breaks all invariance properties (see also Sect.~\ref{S5}).

In latter decades, and also in part due to the impulse given since 1994 by the \emph{International Meeting on Thermodifussion} (IMT) series of conferences~\cite{IMT12_Topical}, research in thermodiffusion has progressed significantly. Nowadays, it can be safely stated that the measurement of Soret coefficients in binary mixtures has become routine, and that the various alternative experimental techniques used by different groups have progressed to a level where agreement is readily reached, and the experimental dataset of reliable $S_T$ values for different binary mixtures has increased steadily~\cite{IMT12_Topical}. Maybe for these reasons the interest of the community has moved towards ternary mixtures in the last years, as a first step towards truly multi-component mixtures. Also, the establishment of large international collaborations associated with space science\footnote{DCMIX (Diffusion Coefficients in ternary MIXtures) and SCCO-SJ10 (Soret Coefficients for Crude Oil at ShiJian-10) are space missions sponsored by the European Space Agency (ESA) in collaboration with Roskosmos and the Chinese Space Administration (CSA), respectively.}, like DCMIX~\cite{MialdunEtAl} or SCCO-SJ10~\cite{miSCCO1,miSCCO2}, has undoubtedly helped to switch the community focus towards ternaries.

\section{Soret coefficients in ternary mixtures\label{S2}}

In a ternary mixture there are two independent concentrations, $x_1$ and $x_2$, while $x_3=1-x_1-x_2$ is a dependent concentration. As a consequence, one initially needs two independent Soret coefficients $S_{T,1}$ and $S_{T,2}$ to describe thermodiffusion in these systems.

Note that the step from binary to ternary systems is not trivial and, for diffusion, requires the introduction of a Fick diffusion matrix $\mathsf{D}$ instead of the single scalar Fick diffusion coefficient $D$ needed for a binary mixture~\cite{TaylorKrishna}. Associated with the matrix description of diffusion there appear several theoretical complications, like the frame-dependence of diffusion matrices, that are difficult to grasp experimentally. Indeed, very few experimental papers in ternary or multi-component mixtures contain any reference to the frame in which their diffusion matrices are measured.

For thermodiffuion and the definition of Soret coefficients there have been some confusion regarding the concentration prefactor~\cite{CostesequeEtAl}. For instance, Shevtsova \emph{et al.}~\cite{Shevtsova_2011} adapted for ternaries a proposal by Kempers~\cite{Kempers98} and used a prefactor $x_1(1-x_1)$ for the first Soret coefficient, and a prefactor $x_2(1-x_2)$ for the second. On the other hand, Ghorayed and Firozabaadi~\cite{GhorayedFiroo} proposed to use prefactors $x_1x_3$ for the first Soret and $x_2x_3$ for the second. Because of these inconsistencies, in the first experimental benchmark on thermodiffusion in ternary systems~\cite{BenchmarkBayonne} it was decided to introduce new Soret coefficients $S_{T,i}^\prime$  simply defined as the ratio between concentration (in mass fraction) and temperature gradient, without any prefactors. A similar approach was adopted when reporting the results of the SCCO-SJ10 space mission~\cite{miSCCO2}. Notice that none of these options retain for ternaries the two properties, enumerated in Sect.~\ref{S1}, which make the definition Eq.~\eqref{E01} of Soret coefficients so convenient for binary mixtures.

The main purpose of this paper is to show how to introduce a concentration prefactor in the definition of Soret coefficients for a ternary mixture that retains all the convenient properties that the $S_T$ of Eq.~\eqref{E01} has for binaries. The inconvenience is that such a prefactor has to be in the form of a matrix. If one defines Soret coefficients for a ternary mixture as:
\begin{equation}\label{E04}
\begin{bmatrix}x_1 (1-x_1) & -x_1x_2\\
-x_1x_2& x_2(1-x_2)
\end{bmatrix}
\begin{pmatrix}
S_{T,1}\\S_{T,2}\end{pmatrix}~\nabla{T}=-
\begin{pmatrix}\nabla{x}_1\\\nabla{x}_2\end{pmatrix},
\end{equation}
the resulting Soret coefficients, $S_{T,1}$ and $S_{T,2}$, are independent of whether concentrations are expressed in mol or mass fraction. Indeed, simple differentiation of the relationship between concentrations in mass and mol fractions shows that (see exercise~1.5 in Ref.~\cite{TaylorKrishna}):
\begin{multline}\label{E05}
\begin{bmatrix}x_1 (1-x_1) & -x_1x_2\\
-x_1x_2& x_2(1-x_2)
\end{bmatrix}^{-1}
\begin{pmatrix}\nabla{x}_1\\\nabla{x}_2\end{pmatrix}\\ =
\begin{bmatrix}w_1 (1-w_1) & -w_1w_2\\
-w_1w_2& w_2(1-w_2)
\end{bmatrix}^{-1}
\begin{pmatrix}\nabla{w}_1\\\nabla{w}_2\end{pmatrix},
\end{multline}
similarly to Eq.~\eqref{E02}. Note that the definition of Soret coefficients given by Eq.~\eqref{E04} is not only frame-invariant\footnote{Al least in the mol and mass reference frames. Some considerations about the volume frame of reference follow in Sect.~\ref{S5}}, but it also cancels the two concentration gradients at all pure component limits which, for a ternary mixture are three: Pure component-1, $x_1=1$ and, thus, $x_2=0$. Pure component-2, $x_2=1$ and, thus, $x_1=0$. Pure component-3, $x_1=0$ and $x_2=0$.

Although the question of the binary limits of the ternary concentration triangle will be detailedly addressed later in Sect.~\ref{S2B}, one direct consequence of Eq.~\eqref{E04} is that, when $x_1\to0$ or $x_2\to0$, one of the Soret coefficients will approach to the Soret coefficient measured in the corresponding binary mixture, with the correct concentration prefactor as given by Eqs.~\eqref{E01} or~\eqref{E03}. When one of the concentrations goes to zero, there is no need of conversion factors to compare  Soret coefficients measured in a ternary mixture with binary tabulated values of $S_T$.

As already mentioned, one disadvantage of Eq.~\eqref{E04} is the matrix character of the concentration prefactor, implying that Soret coefficient \#1 will not only depend on the ratio of $\nabla{x}_1/\nabla{T}$, but on a linear combination of $\nabla{x}_1/\nabla{T}$ and $\nabla{x}_2/\nabla{T}$. Hence, in general, no Soret coefficient can be unequivocally assigned to each one of the individual components of the mixture. Although in Sect.~\ref{S2B}, when discussing binary limits more deeply, the physical interpretation of $S_{T,i}$ will be revisited, the mix-up of components implied by Eq.~\eqref{E04} is not fully foreign to transport in ternary mixtures. For instance, it is becoming common to report experimental values for the eigenvalues of the diffusion matrix~\cite{BenchmarkBayonne}, which, as the Soret coefficients of Eq.~\eqref{E04}, cannot be unequivocally assigned to individual components of the mixture.

The presentation of Eqs.~\eqref{E04} and~\eqref{E05} is the main purpose of the present work. In the remainder of this paper the contents of Sect.~\ref{S2} are complemented with additional, somewhat more technical, discussions on the general expression of diffusion fluxes, binary limits, selection of dependent concentration, the usage of molar concentrations and a possible extension to multi-component mixtures.

\section{General expression of diffusion fluxes}

The contents of Sects.~\ref{S1} and~\ref{S2} above refer to a steady-state, in which the total mass fluxes (containing both purely diffusive and thermodiffusive contributions) vanish. However, as is the case for binary mixtures, the definition~\eqref{E04} of Soret coefficients for a ternary mixture can also be extended to transient situations, where mass fluxes are not zero. Then, as extensively discussed elsewhere~\cite{TaylorKrishna}, one has to distinguish between frames of reference. Let us consider first the mol-average frame of reference. Inspired by Eq.~\eqref{E04} we write, in general, the mol diffusion fluxes relative to the mol average velocity, $J_i^\text{(x)}$ (units of mol~m$^{-2}$s$^{-1}$), as:
\begin{equation}\label{E06}\begin{pmatrix}{J}^\text{(x)}_1\\{J}^\text{(x)}_2\end{pmatrix}=
-c_t~\mathsf{D}^\text{(x)} \cdot \left\{\begin{pmatrix}\nabla{x}_1\\\nabla{x}_2\end{pmatrix}+\mathsf{X}\cdot
\begin{pmatrix}
S_{T,1}\\S_{T,2}\end{pmatrix}~\nabla{T}\right\},
\end{equation}
where $\mathsf{X}$ represents a shorthand for the matrix
\begin{equation}\label{E07}
\mathsf{X}=\begin{bmatrix}x_1 (1-x_1) & -x_1x_2\\
-x_1x_2& x_2(1-x_2)
\end{bmatrix},
\end{equation}
$\mathsf{D}^\text{(x)}$ is the Fick diffusion matrix in the mol frame of reference and $c_t=c_1+c_2+c_3$ is the total molar density of the mixture. Next, using Eq.~\eqref{E05}, one can rewrite Eq.~\eqref{E06} as
\begin{equation*}\begin{pmatrix}{J}^\text{(x)}_1\\{J}^\text{(x)}_2\end{pmatrix}=
-c_t~\mathsf{D}^\text{(x)} \cdot \mathsf{X}\cdot\mathsf{W}^{-1} \left\{\begin{pmatrix}\nabla{w}_1\\\nabla{w}_2\end{pmatrix}+\mathsf{W}\cdot
\begin{pmatrix}
S_{T,1}\\S_{T,2}\end{pmatrix}~\nabla{T}\right\},
\end{equation*}
where the matrix $\mathsf{W}$ is similar to $\mathsf{X}$ in Eq.~\eqref{E07}, but with the concentrations expressed in mass fractions. Next, recalling~\cite{TaylorKrishna} the relationship between the mol diffusion fluxes relative to the mol average velocity, $J_i^\text{(x)}$, and the mass diffusion fluxes relative to the center of mass velocity, $J_i^\text{(w)}$ (units of kg~m$^{-2}$s$^{-1}$)
\begin{equation*}\begin{pmatrix}{J}^\text{(w)}_1\\{J}^\text{(w)}_2\end{pmatrix}=
\frac{\rho_t}{c_t}~\mathsf{W}\cdot\mathsf{X}^{-1} \cdot \begin{pmatrix}{J}^\text{(x)}_1\\{J}^\text{(x)}_2\end{pmatrix},
\end{equation*}
it can be shown that
\begin{equation}\label{E08}\begin{pmatrix}{J}^\text{(w)}_1\\{J}^\text{(w)}_2\end{pmatrix}=
-\rho_t~\mathsf{D}^\text{(w)} \cdot \left\{\begin{pmatrix}\nabla{w}_1\\\nabla{w}_2\end{pmatrix}+\mathsf{W}\cdot
\begin{pmatrix}
S_{T,1}\\S_{T,2}\end{pmatrix}~\nabla{T}\right\},
\end{equation}
where $\rho_t=\rho_1+\rho_2+\rho_3$ is the total mass density, and where
\begin{equation}\label{E09}
\mathsf{W}^{-1}\cdot\mathsf{D}^\text{(w)}\cdot\mathsf{W}=\mathsf{X}^{-1}\cdot\mathsf{D}^\text{(x)} \cdot \mathsf{X}.
\end{equation}
Of course, Eq.~\eqref{E09} is just the standard relationship between Fick diffusion matrices in the mass and in the mol frame of references and, taking into account the different notations, it is exactly the same as Eq.~(3.2.11) in the book by Taylor and Krishna~\cite{TaylorKrishna}.

On comparing Eq.~\eqref{E06} with Eq.~\eqref{E08} it is concluded that the use of concentration prefactors, $\mathsf{X}$ or $\mathsf{W}$, makes the Soret coefficients of Eq.~\eqref{E04} frame invariant\footnote{Al least in the mol and mass reference frames. Some considerations about the volume frame of reference follow in Sect.~\ref{S5}} in the general case, not only in the steady state. That is, in the non-isothermal case, Soret coefficients can be defined so that all the frame-dependence is carried on by the diffusion matrix only, which appears as an overall multiplicative prefactor in either Eq.~\eqref{E06} or Eq.~\eqref{E08}, being the same matrix measured in isothermal conditions. The contribution of thermodiffusion to the total diffusive fluxes, even when they are not zero, can be described in a frame-independent manner.

The use of Eq.~\eqref{E09} to express the relation between Fick diffusion matrices in different reference frames represents a simplification when compared to the traditional way of expressing it~\cite{TaylorKrishna}. First of all, current Eq.~\eqref{E09} is more compact, it requires less matrices. In addition, Eq.~\eqref{E09} clearly shows that in binary systems, where all matrices reduce to scalars, diffusion is frame-independent; while in the ternary case is not. It also gives indications towards a possible frame-independent description of isothermal diffusion in ternaries, although this research will not be further pursued here.

\section{Binary limits\label{S2B}}

As often in ternary mixtures, it is interesting to discuss Eq.~\eqref{E04} in the three binary limits of the ternary concentration triangle. Initially, one has two degrees of freedom for the specification of the coefficients $S_{T,1}$ and $S_{T,2}$. However, consistency with Soret coefficients measured in the corresponding binary mixture, only leaves one degree of freedom over the three sides of the concentration triangle. In particular, the following connections with the binary Soret coefficients must hold:
\begin{subequations}\label{E10}
\begin{align}
\text{When~}&x_1\to 0,& S_{T,2}&\to S^{(3,\text{b})}_{T,2}\label{E10A}\\
\text{When~}&x_2\to 0,& S_{T,1}&\to S^{(3,\text{b})}_{T,1}\label{E10B}\\
\text{When~}&x_3\to 0,& S_{T,1}-S_{T,2}&\to S^{(2,\text{b})}_{T,1}\label{E10C}
\end{align}
\end{subequations}
where $S^{({j,\text{b}})}_{T,i}$ denotes the corresponding Soret coefficient measured in the binary mixture of $i$ and $j$, with $j$ being the dependent concentration. As anticipated before, comparison with Soret coefficients measured in binary mixtures is direct and does not require the use of any conversion or data manipulation.

Moreover, Eqs.~\eqref{E10} also provide a clue for a possible physical interpretation of the Soret coefficients $S_{T,1}$ and $S_{T,2}$ introduced by Eq.~\eqref{E04}. One can imagine that Eq.~\eqref{E04} represents thermodiffusion of 1 and 2 in the presence of 3 as a linear combination of (1,3) thermodiffusion with (2,3) thermodiffusion; with $S_{T,1}$ representing thermodiffusion of 1 in 3, congruent with Eq.~\eqref{E10B}; and $S_{T,2}$ representing thermodiffusion of 2 in 3, congruent with Eq.~\eqref{E10A}. The congruence of Eq.~\eqref{E10C} within this picture will be analyzed next.

\section{Choice of dependent concentration\label{S3}}

In a binary mixture one has two choices for the dependent concentration and, thus, two possible Soret coefficients to describe thermodiffusion. However, according to Eq.~\eqref{E01}, one of these Soret coefficients is  minus the other, so that switching between them is quite straightforward.

In a ternary mixture the situation is more complicated since one has three choices for the dependent concentration, meaning three different possible pairs of Soret coefficients. However, only one of these pairs is independent. To establish the relationships between these different Soret pairs one uses $x_1+x_2+x_3=1$ to deduce from Eq.~\eqref{E04}:
\begin{equation*}
\begin{bmatrix}x_1 (1-x_1) & -x_1x_3\\
-x_1x_3& x_3(1-x_3)
\end{bmatrix}\normalsize
\begin{bmatrix}
S_{T,1}-S_{T,2}\\-S_{T,2}\end{bmatrix}\nabla{T}=-
\begin{pmatrix}\nabla{x}_1\\\nabla{x}_3\end{pmatrix},
\end{equation*}
and
\begin{equation*}
\begin{bmatrix}x_2 (1-x_2) & -x_2x_3\\
-x_2x_3& x_3(1-x_3)
\end{bmatrix}
\begin{bmatrix}
S_{T,2}-S_{T,1}\\-S_{T,1}\end{bmatrix}\nabla{T}=-
\begin{pmatrix}\nabla{x}_2\\\nabla{x}_3\end{pmatrix}.
\end{equation*}
To summarize the expressions above it is convenient to adopt the nomenclature $S_{T,i}^{(j)}$ with superscript $j$ designating the component of the ternary mixture whose concentration is considered as dependent variable; and subscript $i$ like in Eq.~\eqref{E04}, associated to the corresponding independent variable. Then, for instance, the two independent Soret coefficients used so far become $S_{T,1}^{(3)}$ (associated to component 1) and $S_{T,2}^{(3)}$ (associated to component 2). Adopting this nomenclature, the two precedent equations imply:
\begin{equation}\label{E11}
\begin{aligned}
S_{T,1}^{(2)}&=S_{T,1}^{(3)}-S_{T,2}^{(3)},&S_{T,3}^{(2)}&=-S_{T,2}^{(3)},\\
S_{T,2}^{(1)}&=S_{T,2}^{(3)}-S_{T,1}^{(3)},&S_{T,3}^{(1)}&=-S_{T,1}^{(3)}.
\end{aligned}
\end{equation}
The four Eqs.~\eqref{E11} mean that, of the initially three different pairs of Soret coefficients, only one pair is independent. It is also interesting to note the consistency with Eq.~\eqref{E10C} and with the situation in binaries, where exchange of independent concentration implies a sign change in the Soret coefficient.

Since there are more Soret coefficients, the relationships among them, Eqs.~\eqref{E11}, are a bit more complicated than in the case of binaries. However, simple `circular' relations like
\begin{equation}\label{E12}
\begin{aligned}
S_{T,1}^{(3)}+S_{T,3}^{(2)}+S_{T,2}^{(1)}&=0,\\
S_{T,2}^{(3)}+S_{T,3}^{(1)}+S_{T,1}^{(2)}&=0,
\end{aligned}
\end{equation}
hold, which would be certainly very useful when changing the order of components. Alternative proposals to describe the Soret effect in ternary mixtures using thermal diffusion ratios~\cite{GebhardtKoehler15}, cite as an advantage the existence of relationships similar to the current Eq.~\eqref{E12}.

\section{Molar concentrations\label{S5}}

It is also quite common to express the concentrations of components in a mixture as mol per unit volume, $c_i=c_tx_i$. However, because typically $c_t$ depends on temperature, this choice introduces complications when temperature gradients are present, and has not been popular for the description of thermodiffusion. For instance, in the case of a non-isothermal binary mixture, one has that in general $\nabla(c_1+c_2)\neq0$, in contrast with concentrations in mol or mass fraction where $\nabla(x_1+x_2)=\nabla(w_1+w_2)=0$ always.

In spite of these complications, some books or reviews present the theory of thermodiffusion in binaries in terms of molar concentrations~\cite{KjelstrupBook2010,PiazzaParola}, as well as some experimentalists report binary Soret coefficients using these concentration units~\cite{MastBraun,Matsuura15}. However, a detailed examination shows that in these works it is implicitly assumed: \emph{(i)} That the dependence of $c_t$ on temperature can be neglected. \emph{(ii)} That the concentration $x$ of the `solute' is very low, so that $1-x\simeq 1$. In that case, multiplying Eq.~\eqref{E01} for a binary mixture by $c_t$, one obtains~\cite{KjelstrupBook2010,PiazzaParola}:
\begin{equation}\label{E13}
cS_T~\nabla{T} = -\nabla{c}.
\end{equation}
Equation~\eqref{E13} shows that, with the restrictions mentioned above, Soret coefficients can be defined in a binary mixture using molar concentrations, and that they are equal to the more general definitions of Eq.~\eqref{E01} or Eq.~\eqref{E03}, in terms of mol fraction or mass fraction, respectively.

Solving Eq.~\eqref{E13} leads to the so-called exponential depletion law~\cite{DuhrBraun}, which is the starting point of many experimental works reporting Soret coefficients, in particular of macromolecules, biomolecules or colloidal particles~\cite{MastBraun,Matsuura15,DuhrBraun,IacopiniEtAl}. In these cases the assumptions mentioned above typically hold. It is also customary to use the term thermophoresis to refer to thermodiffusion in these colloidal or macromolecular systems with low `solvent' concentration.

Regarding ternary mixtures it is not the intention of this paper to go into the details,  but one can convince oneself readily that, within the same approximations mentioned above for binary mixtures, Soret coefficients can be defined on the basis of molar concentrations that will be numerically the same as those of Eq.~\eqref{E04}.

\section{Extension to multi-component mixtures}

This work has been so far restricted to a explicit discussion for ternary mixtures, mainly because of the recent experimental interest on thermodiffusion in these systems~\cite{MialdunEtAl,miSCCO1,miSCCO2}. However, the ideas  presented here can be generalized to multi-component mixtures. Indeed, for a $n$-component mixture one can define $(n-1)\times(n-1)$ matrices $\mathsf{X}$ and $\mathsf{W}$ as:
\begin{align}
X_{ij}&=x_i\delta_{ij}-x_ix_j,&W_{ij}&=w_i\delta_{ij}-w_iw_j,
\end{align}
with $i,j\in\{1,n-1\}$ being the independent components, while $x_n$ or $w_n$ are the dependent concentrations and $\delta_{ij}$ the Kronecker delta. Then, a little bit of effort shows that Eq.~\eqref{E05} holds in general, namely:
\begin{equation}
\mathsf{X}^{-1} \cdot \begin{bmatrix}\nabla{x}_1\\\nabla{x}_2\\\vdots\\\nabla{x}_{n-1}\end{bmatrix} = \mathsf{W}^{-1}\cdot  \begin{bmatrix}\nabla{w}_1\\\nabla{w}_2\\\vdots\\\nabla{w}_{n-1}\end{bmatrix}.
\end{equation}
Hence, matrices $\mathsf{X}$ and $\mathsf{W}$ can be used to define frame-independent Soret coefficients for multi-component systems in an exactly similar fashion as explicitly elucidated in this paper ternaries. In particular, Eqs.~\eqref{E06}, \eqref{E08} and~\eqref{E09} hold in general, for an arbitrary number of components, which opens the way to a possible description of thermodiffusion in a frame-independent way.

\section*{Acknowledgements}

This research has been in part supported by grant ESP2017-83544-C3-2-P of the Spanish \emph{Agencia Estatal de Investigaci\'on}. The author has greatly benefitted from his participation in the Scientific Committees of the ESA missions DCMIX and SCCO-SJ10. My most sincere acknowledgement to all co-members of these committees, discussions with whom are at the origin of the present work. Very useful advice from Prof. Jan V. Sengers is also warmly acknowledged.

%

\begin{thebibliography}{10}
\providecommand{\url}[1]{{#1}}
\providecommand{\urlprefix}{URL }
\expandafter\ifx\csname urlstyle\endcsname\relax
  \providecommand{\doi}[1]{DOI \discretionary{}{}{}#1}\else
  \providecommand{\doi}{DOI \discretionary{}{}{}\begingroup
  \urlstyle{rm}\Url}\fi

\bibitem{DeGrootMazur}
S.R. {de Groot}, P.~Mazur, \emph{Non-Equilibrium Thermodynamics}
  (North-Holland, Amsterdam, 1962).
\newblock Dover edition, 1984

\bibitem{Demirel}
Y.~Demirel, \emph{Nonequilibrium Thermodynamics} (Elsevier, Amsterdam, 2002)

\bibitem{LebonJouCasas}
G.~Lebon, D.~Jou, J.~{Casas-V\'azquez}, \emph{Understanding Nonequilibrium
  Thermodynamics} (Springer, 2008)

\bibitem{KjelstrupBook2010}
S.~Kjelstrup, D.~Bedeaux, E.~Johannessen, J.~Gross, \emph{Non-Equilibrium
  Thermodynamics for Engineers} (World Scientific, Singapore, 2010)

\bibitem{PlattenCosteseque}
J.K. Platten, P.~Costes{\`e}que, Eur. Phys. J. E \textbf{15}, 235 (2004)

\bibitem{BeltonTyrrell}
P.S. Belton, H.J.V. Tyrrell, Z. Naturforsch. \textbf{26}, 48 (1971)

\bibitem{WiegandKitaNing}
S.~Wiegand, H.~Ning, R.~Kita, J. Non-Equilib. Thermodyn. \textbf{32}, 193
  (2007).

\bibitem{CabreraEtAl}
H.~Cabrera, F.~Cordido, A.~Vel\'asquez, P.~Moreno, E.~Sira, S.A.
  L\'opez-Rivera, C.R. Mecanique \textbf{341}, 372 (2013).

\bibitem{Hafskjold17}
B.~Hafskjold, Eur. Phys. J. E \textbf{40}, 4 (2017)

\bibitem{DiLecce17}
S.~{Di Lecce}, T.~Albrecht, F.~Bresme, Scientific Reports \textbf{7}, 44833
  (2017)

\bibitem{KohlerMuller95}
W.~K\"ohler, B.~M\"uller, J. Chem. Phys. \textbf{103}, 4367 (1995)

\bibitem{MialdunShevtsova}
A.~Mialdun, V.M. Shevtsova, Int. J. Heat Mass Trans. \textbf{51}, 3164 (2008).

\bibitem{FurtadoEtAl}
F.A. Furtado, A.J.S. ~, C.R.A. Abreu, F.W. Tavares, Brazil. J. Chem. Eng.
  \textbf{32}, 683 (2015).

\bibitem{DeGroot1942}
S.R. De~Groot, Physica \textbf{9}(7), 699 (1942).

\bibitem{IMT12_Topical}
R.~Delgado-Buscalioni, M.~Khayet, J.M. {Ortiz de Z\'arate}, F.~Croccolo, Eur.
  Phys. J. E \textbf{40}, 51 (2017).

\bibitem{MialdunEtAl}
A.~Mialdun, C.~Minetti, Y.~Gaponenko, V.~Shevtsova, F.~Dubois, Microgravity
  Sci. Technol. \textbf{25}, 83 (2013)

\bibitem{miSCCO1}
G.~Galliero, H.~Bataller, F.~Croccolo, R.~Vermorel, P.~Artola, B.~Rousseau,
  V.~Vesovic, M.~{Bou-Ali}, J.M. {Ortiz de Z{\'a}rate}, K.~Zhang, F.~Montel,
  Microgravity Sci. Technol. \textbf{28}, 79 (2016)

\bibitem{miSCCO2}
G.~Galliero, H.~Bataller, J.P. Bazile, J.~Diaz, F.~Croccolo, H.~Hoang,
  R.~Vermorel, P.A. Artola, B.~Rousseau, V.~Vesovic, M.M. Bou-Ali, J.M. {Ortiz
  de Z\'arate}, S.~Xu, K.~Zhang, F.~Montel, A.~Verga, O.~Minster, npj
  microgravity \textbf{3}, 20 (2017).

\bibitem{TaylorKrishna}
R.~Taylor, R.~Krishna, \emph{Multicomponent Mass Transfer} (Wiley, New York,
  1993)

\bibitem{CostesequeEtAl}
P.~Costes{\`e}que, A.~Mojtabi, J.K. Platten, C. R. Mecanique \textbf{339}, 275
  (2011).

\bibitem{Shevtsova_2011}
V.~Shevtsova, V.~Sechenyh, A.~Nepomnyashchy, J.C. Legros, Phil. Mag.
  \textbf{91}(26), 3498 (2011).

\bibitem{Kempers98}
L.J.T.M. Kempers, J. Chem. Phys. \textbf{90}, 6541 (1998).

\bibitem{GhorayedFiroo}
K.~Ghorayeb, A.~Firoozabadi, AIChe J. \textbf{46}, 883 (2000).

\bibitem{BenchmarkBayonne}
M.M. Bou-Ali, A.~Ahadi, D.~{Alonso de Mezquia}, Q.~Galand, M.~Gebhardt,
  O.~Khlybov, W.~K\"{o}hler, M.L. naga, J.C. Legros, T.~Lyubimova, A.~Mialdun,
  I.~Ryzhkov, M.Z. Saghir, V.~Shevtsova, S.V. Vaerenbergh, Eur. Phys. J. E
  \textbf{38}, 30 (2015).

\bibitem{GebhardtKoehler15}
M.~Gebhardt, W.~K\"{o}hler, J. Chem. Phys. \textbf{143}, 164511 (2015).

\bibitem{PiazzaParola}
R.~Piazza, A.~Parola, J. Phys.: Condens. Matter \textbf{20}, 153102 (2008).
\newblock \urlprefix\url{http://stacks.iop.org/JPhysCM/20/153102}

\bibitem{MastBraun}
C.B. Mast, D.~Braun, Phys. Rev. Lett. \textbf{104}, 188102 (2010)

\bibitem{Matsuura15}
H.~Matsuura, S.~Iwaasa, Y.~Nagasaka, J. Chem. Eng. Data \textbf{60}, 3621–3630
  (2015).

\bibitem{DuhrBraun}
S.~Duhr, D.~Braun, PNAS \textbf{26}, 19678 (2006)

\bibitem{IacopiniEtAl}
S.~Iacopini, R.~Rusconi, R.~Piazza, Eur. Phys. J. E \textbf{19}, 59 (2006)

\end{thebibliography}

\end{document}